\documentclass[aps,prl,byrevtex,showpacs,superscriptaddress,twocolumn,letter]{revtex4-1}

\usepackage{graphicx}
\usepackage{dcolumn}
\usepackage{amsmath}
\usepackage{epsfig}
\usepackage{bm}
\usepackage{amssymb}

\begin{document}

\title{Five measurement bases determine pure quantum states on any dimension}

\author{D.~Goyeneche}
\email{dardo.goyeneche@cefop.udec.cl}
\affiliation{Departamento de F\'{i}sica, Universidad de Concepci\'{o}n, Casilla 160-C, Concepci\'{o}n, Chile}
\affiliation{Center for Optics and Photonics, Universidad de Concepci\'{o}n, Casilla 4016, Concepci\'{o}n, Chile}
\affiliation{MSI-Nucleus on Advanced Optics, Universidad de Concepci\'on, Casilla 160-C, Concepci\'on, Chile}
\author{G.~Ca\~nas}
\affiliation{Departamento de F\'{i}sica, Universidad de Concepci\'{o}n, Casilla 160-C, Concepci\'{o}n, Chile}
\affiliation{Center for Optics and Photonics, Universidad de Concepci\'{o}n, Casilla 4016, Concepci\'{o}n, Chile}
\affiliation{MSI-Nucleus on Advanced Optics, Universidad de Concepci\'on, Casilla 160-C, Concepci\'on, Chile}
\author{S.~Etcheverry}
\affiliation{Departamento de F\'{i}sica, Universidad de Concepci\'{o}n, Casilla 160-C, Concepci\'{o}n, Chile}
\affiliation{Center for Optics and Photonics, Universidad de Concepci\'{o}n, Casilla 4016, Concepci\'{o}n, Chile}
\affiliation{MSI-Nucleus on Advanced Optics, Universidad de Concepci\'on, Casilla 160-C, Concepci\'on, Chile}
\author{E.~S.~G\'omez}
\affiliation{Departamento de F\'{i}sica, Universidad de Concepci\'{o}n, Casilla 160-C, Concepci\'{o}n, Chile}
\affiliation{Center for Optics and Photonics, Universidad de Concepci\'{o}n, Casilla 4016, Concepci\'{o}n, Chile}
\affiliation{MSI-Nucleus on Advanced Optics, Universidad de Concepci\'on, Casilla 160-C, Concepci\'on, Chile}
\author{G.~B.~Xavier}
\affiliation{Center for Optics and Photonics, Universidad de Concepci\'{o}n, Casilla 4016, Concepci\'{o}n, Chile}
\affiliation{MSI-Nucleus on Advanced Optics, Universidad de Concepci\'on, Casilla 160-C, Concepci\'on, Chile}
\affiliation{Departamento de Ingenier\'ia El\'ectrica, Universidad de Concepci\'on, 160-C Concepci\'on, Chile}
\author{G. Lima}
\affiliation{Departamento de F\'{i}sica, Universidad de Concepci\'{o}n, Casilla 160-C, Concepci\'{o}n, Chile}
\affiliation{Center for Optics and Photonics, Universidad de Concepci\'{o}n, Casilla 4016, Concepci\'{o}n, Chile}
\affiliation{MSI-Nucleus on Advanced Optics, Universidad de Concepci\'on, Casilla 160-C, Concepci\'on, Chile}
\author{A. Delgado}
\affiliation{Departamento de F\'{i}sica, Universidad de Concepci\'{o}n, Casilla 160-C, Concepci\'{o}n, Chile}
\affiliation{Center for Optics and Photonics, Universidad de Concepci\'{o}n, Casilla 4016, Concepci\'{o}n, Chile}
\affiliation{MSI-Nucleus on Advanced Optics, Universidad de Concepci\'on, Casilla 160-C, Concepci\'on, Chile}

\begin{abstract}

A long standing problem in quantum mechanics is the minimum number of observables required for the characterisation of unknown pure quantum states. The solution to this problem is specially important for the developing field of high-dimensional quantum information processing. In this work we demonstrate that any pure $d$-dimensional state is unambiguously reconstructed by measuring 5 observables, that is, via projective measurements onto the states of 5 orthonormal bases. Thus, in our method the total number of different measurement outcomes ($5d$) scales linearly with $d$. The state reconstruction is robust against experimental errors and requires simple post-processing, regardless of $d$. We experimentally demonstrate the feasibility of our scheme through the reconstruction of 8-dimensional quantum states, encoded in the momentum of single photons.

\end{abstract}

\maketitle
With the development of high-dimensional quantum information processing techniques \cite{Monz, Lima13, Fabio14, Krenn,Boyd}, the total dimension $d$ of quantum systems employed in experiments increases at a fast pace. Since the total number $\cal{M}$ of measurement outcomes required by conventional quantum tomography methods scales with $d^2$ \cite{DAriano1,DAriano2,James,Thew,Prugovecki,Flammia,Renes,Kaznady,Hradil1,Hradil2,Blume,Maciel,Rau}, it is of paramount importance the search for tomographic protocols specially adapted to higher dimensions. That is, schemes which require a lower $\cal{M}$ and a reduced complexity of the post-processing methods. Thus, it is possible to consider \textit{a priori} information about the set of states to be characterised. For example, rank-$r$ quantum states are reconstructed, with high probability, with $\cal{M}$ of the order of $rd(\log d)^2$ via compressed sensing techniques \cite{Gross}. Nearly matrix product states are determined with $\cal{M}$ linear in $d$ and post-processing that is polynomial in the system size \cite{Cramer}. Permutationally invariant quantum states of $n$ qubits ($d=2^n$) are reconstructed with $\mathcal{M}=(log(d)+2)d$ \cite{Klimov}.

In this work we study the characterisation of unknown pure quantum states via projective measurements. In 1933, W. Pauli \cite{Pauli} considered the unambiguous characterisation of pure states from probability distributions generated by the measurement of a fixed set of observables. It has been shown that the number of observables must be larger than $3$ for $d\ge9$ \cite{Moroz}. Here, we show that almost all pure quantum states can be characterised by a set of probability distributions generated by 4 observables, that is, projective measurements onto a fixed set of $4$ bases $\mathcal{B}_i$ ($i=1,\dots,4$) independently of the dimension $d$. In our case $\mathcal{M}=4d$, which is an improvement compared with quantum tomography based on mutually unbiased bases \cite{Prugovecki}, SIC-POVM \cite{Flammia,Renes} and compressed sensing \cite{Gross}, which require $\mathcal{M}=d(d+1)$, $\mathcal{M}=d^2$ and $\mathcal{M}$ of the order of $d(\log d)^2$, respectively. Pure states can also be reconstructed via the expectation values \cite{Heinossari} of a fixed set of observables. In this case, $4d-\alpha$ observables are required at least, where $\alpha$ is a quantity that scales with the logarithm of the dimension. Consequently, the value of $\cal{M}$ is larger than in our proposal.

The exception to our first characterisation is a statistically unlikely null measure set $\Omega$ of pure states. Such states can be characterised by adding a fifth measurement basis $\mathcal{B}_0$, which detects whether a pure state is in $\Omega$ or not. Thus, it is possible to define a new set of $4$ bases to characterise the state. These bases are similar to $\mathcal{B}_i$ but defined on a lower dimensional subspace. Thereby, the set of pure states is decomposed into a finite number of disjoint sets of states such that each set is reconstructed with a particular set of five bases. Thus, any pure state can be reconstructed with projective measurements on no more than $5$ bases, that is, $\mathcal{M}=5d$. An interesting feature of each set of $5$ bases is that they allow us to certify the initial assumption on the purity of the state to be reconstructed. This class of certification can also be achieved by reconstructing pure states via expectation values \cite{Chen} but at the expense of a much higher $\cal{M}$ than in our proposal.

We demonstrate the feasibility of our tomographic scheme with the experimental characterisation of eight-dimensional quantum states encoded on the transverse momentum of single-photons. The preparation of the states is affected by white noise, which renders the states slightly impure, and the detection process by poison noise. In spite of these conditions the obtained fidelities are higher than $0.96 \pm 0.03$ with $\mathcal{M}=40$. This result is in agreement with numerical simulations which consider realistic noise levels in the preparation of the unknown state as well as in the detection process. Our work shows that the effective characterisation of unknown high-dimensional pure quantum states by means of a reduced number of measurement outcomes is feasible.


A $d$-dimensional quantum state can be expanded as $\rho=\frac{1}{d}\mathbb{I}+\sqrt{\frac{d-1}{2d}}\sum_{j=1}^{d^2-1}r_jT_j$, where $\mathbb{I}$ is the identity operator, $r_j=Tr(\rho T_j)$ form the generalised Bloch vector and $T_j$ are a traceless, hermitian representation of the generators of the group $SU(d)$ such as the generalised Gell-Mann basis \cite{Georgi}. This consists of $d-1$ diagonal operators $T_\alpha$ and $d(d-1)$ non-diagonal operators $T_{k,m}$ and $\tilde T_{k,m}$. The latter are given by $T_{k,m}=|k\rangle\langle m|\!+\!|m\rangle\langle k|$ and $\tilde T_{k,m}=-i|k\rangle\langle m|\!+\!i|m\rangle\langle k|$, where $0\leq k< m\leq d-1$.

Any pure state $|\Psi\rangle=\sum_{k=0}^{d-1}c_k|k\rangle$ satisfies the following set of $d(d-1)/2$ equations
\begin{equation}
2c_mc_k^*=\mathrm{Tr}(|\Psi\rangle\langle\Psi|T_{m,k})+i\mathrm{Tr}(|\Psi\rangle\langle\Psi|\tilde{T}_{m,k}).
\label{PHASES}
\end{equation} Interestingly, $d-1$ of the above equations univocally characterise a certain dense set of pure states. We choose the set of $d-1$ equations with $m=k+1$ which allows us to solve Eq. (\ref{PHASES}) recursively, up to a null measure set of pure states. In order to calculate the traces of Eq. (\ref{PHASES}) we consider rank-one projective measurements associated to the eigenvectors of $T_{k,k+1}$ and $\tilde{T}_{k,k+1}$ with eigenvalues $\pm1$, that is, $\mathcal{M}=4d$. These $4d$ vectors can be always sorted in four orthogonal bases. For $d\ge3$ these bases are given by
\begin{eqnarray}
\label{BASE1}
\mathcal{B}_1&=&\{|2\nu\rangle\pm|2\nu+1\rangle\},\ \mathcal{B}_3=\{|2\nu+1\rangle\pm|2\nu+2\rangle\},\\
\mathcal{B}_2&=&\{|2\nu\rangle\pm i|2\nu+1\rangle\},\ \mathcal{B}_4=\{|2\nu+1\rangle\pm i|2\nu+2\rangle\}, \nonumber
\end{eqnarray}
where addition of labels is carried out modulo $d$ and $\nu\in[0,(d-2)/2]$. For odd dimensions we considered the integer part of $(d-2)/2$ and every basis is completed with $|d\rangle$. Defining $p_{\pm}^k$ ($\tilde p_{\pm}^k$) as the probability of projecting the state $|\Psi\rangle$ onto the eigenvector of $T_{k,k+1}$ ($\tilde T_{k,k+1}$) with eigenvalue $\pm1$, Eq. (\ref{PHASES}) becomes $2c_{k}c_{k+1}^*=\Lambda_k$, where $\Lambda_k=\sqrt{(d-1)/2d}[(p_{+}^k-p_{-}^k)+i(\tilde{p}_{+}^k-\tilde{p}_{-}^k)]$. Table \ref{Table} associates bases $\mathcal{B}_i$ with the eigenvectors of $T_{k,k+1}$ and $\tilde{T}_{k,k+1}$ with eigenvalues $\pm1$ and probabilities $p_{\pm}^k$ and $\tilde p_{\pm}^k$, respectively.
\begin{table}
\centering
\begin{tabular}{|c|c|c|}
\hspace{0.1cm}
$\mathcal{B}_1$ ($\mathcal{B}_3$)\hspace{0.1cm} & $T_{k,k+1},\,k$ even (odd)& \hspace{0.1cm}$p_{\pm}^k,\,k$ even (odd)\hspace{0.1cm}\\
$\mathcal{B}_2$ ($\mathcal{B}_4$) & $\tilde{T}_{k,k+1},\,k$ even (odd) & $\tilde{p}_{\pm}^k,\,k$ even (odd)\\
\end{tabular}
\caption{Association between the four bases, $su(d)$ generators and transition probabilities.}
\label{Table}
\end{table}
The equations $2c_{k}c_{k+1}^*=\Lambda_k$ can be recursively solved leading to
\begin{equation}\label{solution}
c_k = \left\{
\begin{array}{c l}
 c_0\prod_{i=0}^{k/2-1} \frac{\Lambda^*_{2i+1}}{\Lambda_{2i}} & k>0\, \mbox{ even,}\vspace{0.3cm}\\
 \frac{\Lambda^*_0}{2c_0}\prod_{i=0}^{(k-3)/2} \frac{\Lambda^*_{2i+2}}{\Lambda_{2i+1}} & k>1\, \mbox{ odd,} 
\end{array}
\right.
\end{equation}
where $c_1=\frac{\Lambda_0^*}{2c_0}$ and $c_0$ is determined by normalisation. If one of the coefficients $c_k$ vanishes then the system of equations cannot be recursively solved. However, the remaining equations and the normalisation condition are enough to reconstruct the state. This also holds in the case of two consecutive vanishing coefficients. For two non-consecutive vanishing coefficients the system of equations has infinite solutions. For instance, in $d=4$ the set of equations is $2c_0c_1^*=\Lambda_0$, $2c_1c_2^*=\Lambda_1$, $2c_2c_3^*=\Lambda_2$ and $2c_3c_4^*=\Lambda_3$. If $c_0=c_2=0$ or $c_1=c_3=0$ then all left sides vanish. Thus, there are states that cannot be singled out via $\mathcal{B}_k$. These states form a manifold $\Omega$ of dimension $d-2$ and, consequently, are statistically unlikely. This means that, if pure states are randomly selected then with probability $1$ they would be out of $\Omega$. Thus, a total of $4d$ measurement outcomes, corresponding to projective measurements onto states of bases $\mathcal{B}_k$, are informationally complete on the set of pure states up to the null measure set $\Omega$. 

The introduction of a fifth basis $\mathcal{B}_0$, the canonical basis which is first to be measured, allows us to determine whether a state belongs to the manifold $\Omega$ or not depending on the number of vanishing coefficients detected. If the state is in $\Omega$ and has $m$ null coefficients, then we can reconstruct the state into the subspace associated to the $d-m$ non-vanishing coefficients. This is done with the bases $\mathcal{B}_k$ ($k=1,2,3,4$) but defined for a subspace of dimension $d-m$. Thereby, any pure state can be reconstructed via projective measurements onto, at most, $5$ orthonormal bases or equivalently with $\mathcal{M}\le 5$.

In the $N$ qubits case, separable bases $\mathcal{B}_0$, $\mathcal{B}_1$ and $\mathcal{B}_2$ correspond to the eigenstates of the local operators $\sigma_z^{(1)}\otimes\dots\otimes\sigma_z^{(N)}$, $\sigma_z^{(1)}\otimes\dots\otimes\sigma_z^{(N-1)}\otimes\sigma_x^{(N)}$ and $\sigma_z^{(1)}\otimes\dots\otimes\sigma_z^{(N-1)}\otimes\sigma_y^{(N)}$, respectively. Entangled bases $\mathcal{B}_3$ and $\mathcal{B}_4$ can be mapped onto bases $\mathcal{B}_1$ and $\mathcal{B}_2$ respectively by applying the quantum Fourier transform (QFT)\cite{Coppersmith,Ekert,Josza} twice. This can be optimally implemented with the order of $N^2$ \cite{Cooley} Hadamard and conditional two-qubit phase gates. Thus, our tomographic scheme is reduced to simple local measurements and QFT. The latter has been experimentally realised in photonic qubits \cite{Lanyon}, nuclear magnetic resonance \cite{Weinstein,Vandersypen}, neutral molecules \cite{Hosaka}, superconducting qubits \cite{Mariantoni},  and trapped ions \cite{Schindler,Chiaverini,Ivanov}.

Measurements onto the set of bases allow us to certify the assumed purity of the unknown state. A state $\rho$ is pure if and only if the equation $|\rho_{k,l}|^2= |\rho_{k,k}||\rho_{l,l}|$ holds for every $k,l=0,\dots,d-1$. Remarkably, these conditions for $l=k+1$ are enough to ensure that $\rho$ determines a pure state. Indeed, given that $\rho$ is a quantum state then $\rho=AA^{\dag}$ for a given operator $A$. So, every entry of $\rho$ satisfies $\rho_{k,l}=v_k\cdot v_l$, where $v_k$ is the $k$th column of $A$. If $l=k+1$ the $d-1$ equations $|\rho_{k,k+1}|^2= |\rho_{k,k}||\rho_{k+1,k+1}|$ hold if and only if vectors $v_k$ and $v_{k+1}$ are parallel for every $k=1,\dots,d-1$. Consequently, $\rho$ is pure. The same holds for any set of five bases.

Tomographic schemes reconstruct quantum states in  matrix space. Since quantum states form a null measure set in matrix space, noisy measurement results lead to matrices that do not represent quantum states. To overcome this problem experimental data is post-processed with maximum likelihood estimation \cite{Hradil1}. An important feature of our method is that it delivers a vector of the underlying Hilbert space for any set of noisy probabilities, being the normalisation of this vector the only procedure required to obtain a pure state from noisy probabilities. 

\begin{figure} 
\centering
{\includegraphics[width=0.47\textwidth]{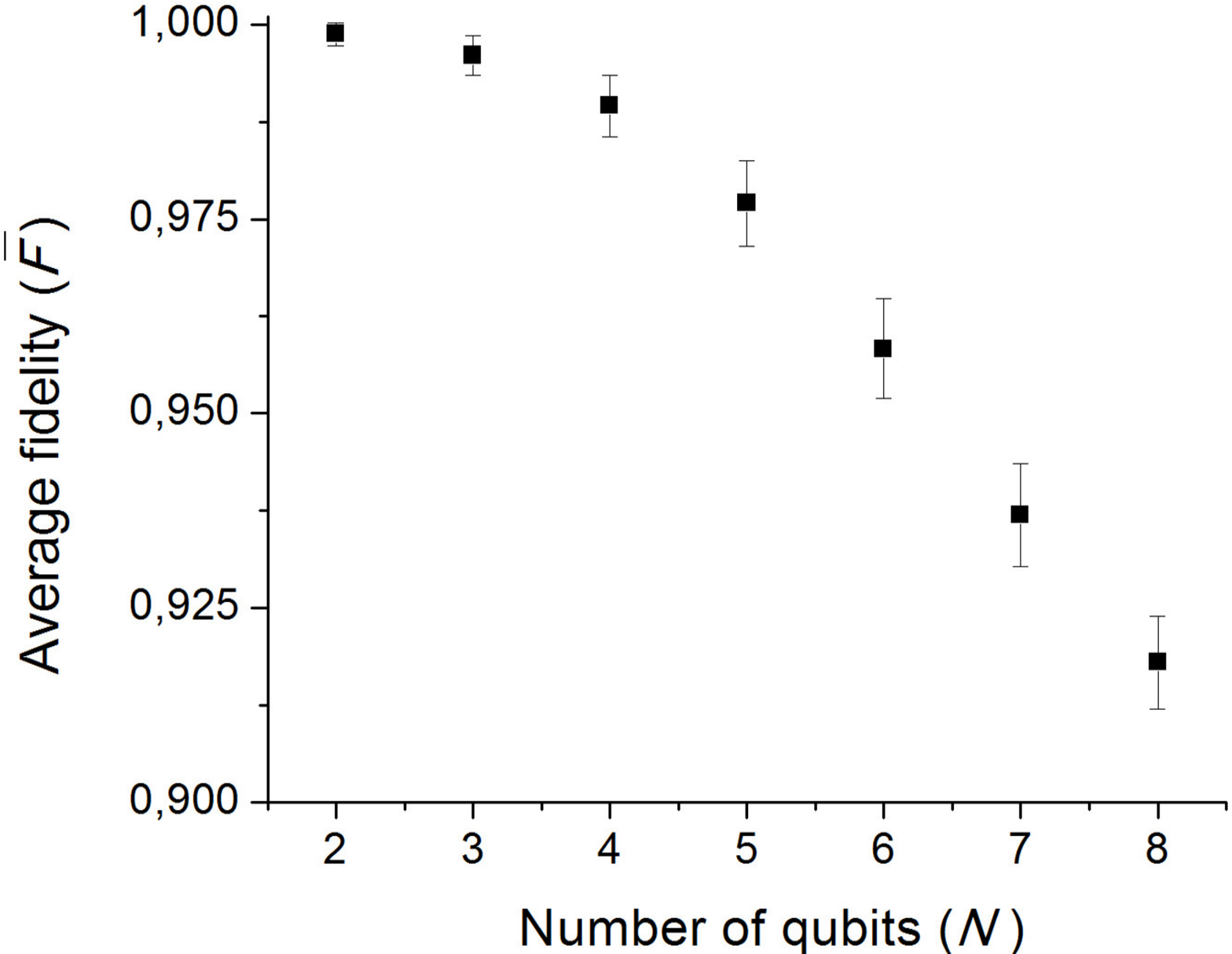}}
\caption{Average fidelity $\bar F$ (black squares) and standard deviation (bars) as a function of the number N of qubits ($d=2^N$). Here we considered a white noise level of $\lambda=0.03$ (see Eq. (\ref{noise-generation})) and poisson noise $\Delta p=0.00081$ in the detection process.}
\label{Fig1}
\end{figure}

The scheme here proposed is based on a priori information about the purity of the state to be reconstructed. This condition is difficult to realise in current experiments. However, it is possible to generate nearly pure states such as
\begin{equation}
\rho=(1-\lambda)|\Psi\rangle\langle\Psi|+\frac{\lambda}{d}\mathbb{I}.
\label{noise-generation}
\end{equation}
Here, the generation process is affected by white noise, which mixes the target pure state $|\Psi\rangle\langle\Psi|$ with the maximally mixed state $\mathbb{I}/d$. The strength of the process is given by the real number $\lambda$. This model is in agreement with our experimental setup. The following lower bound for the fidelity holds 
\begin{equation}
F\ge 1-\Delta p\sqrt{\frac{d-1}{d}}\sum_{k=0}^{d-1}\frac{\sigma_k}{\sigma_k+\sigma_{k+1}},
\label{LowerBound}
\end{equation}
where $\sigma_k=\sqrt{p^0_k-\lambda/d}$, $\{p^0_k\}_{k=0,\dots,d-1}$ is the probability distribution generated by measurements on the canonical basis $\mathcal{B}_0$ and $\Delta p$ is the maximal amount of noise introduced by the photo-detection process. For a uniform state (i.e., $|c_k|^2=1/d$ for every $k=0,\dots,d-1$) of a composite system of $N$ qubits we obtain $F\ge 1-\Delta p2^{N-1}$. Thus, to keep a  constant large lower bound while increasing the number of qubits, $\Delta p$ must decrease as $2^{1-N}$. Fig. \ref{Fig1} shows the average fidelity $\bar F=\int |\langle\Psi|\Psi_{est}\rangle|d\Psi$ onto the complete Hilbert space and the square root of its variance \cite{Note} as functions of $N$ considering white and poisson noises.

\begin{figure}
\centering
\includegraphics[width=0.47\textwidth]{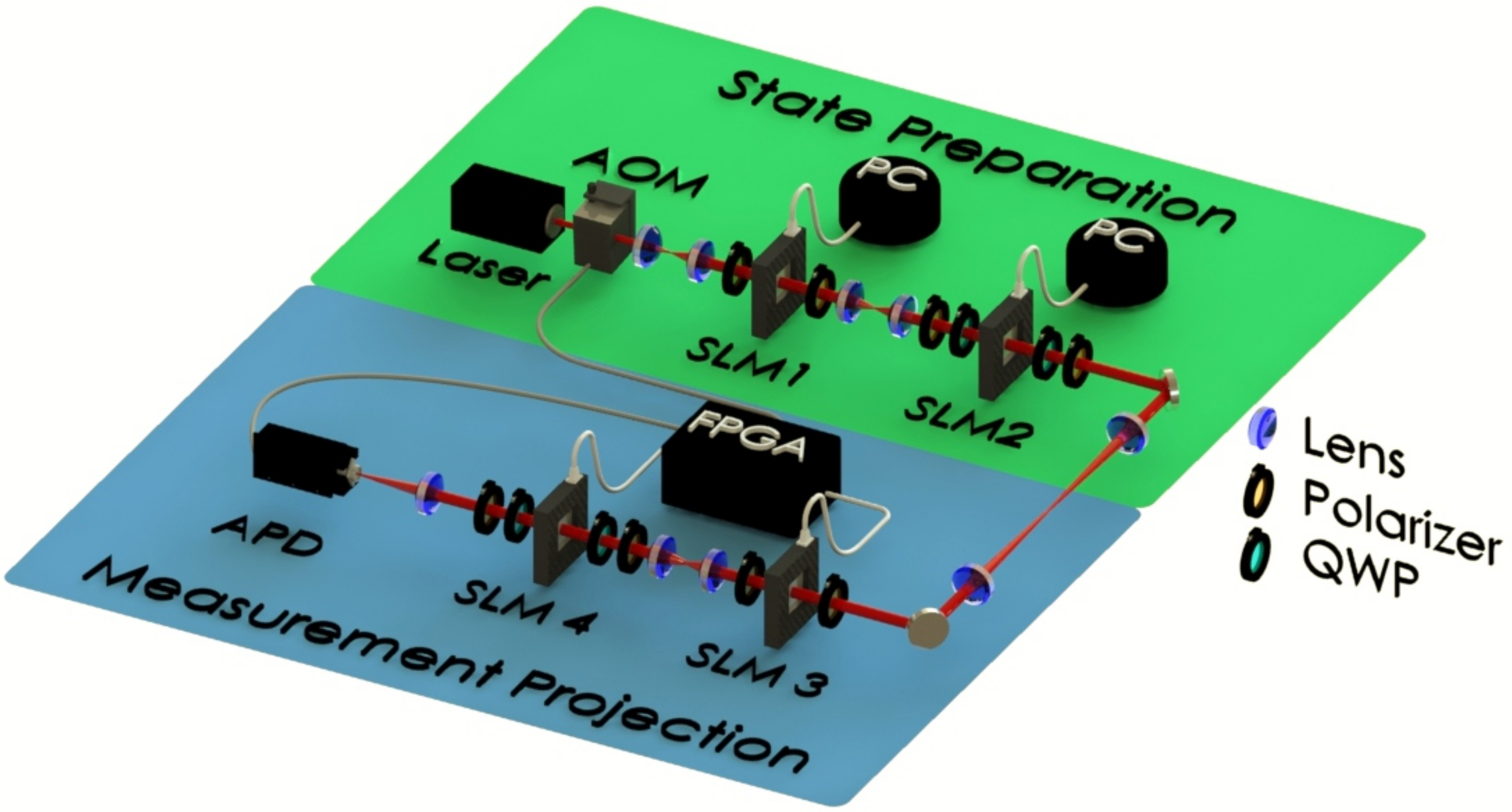}
\caption{State generation and projective measurement stages of the experimental setup. A cw laser and an AOM create single photon pulses which are transformed by computer-controlled (PC) SLM1 and SLM2 into a qudit state. This is projected onto any other qudit state by SLM3 and SLM4 and a point-like avalanche photodetector (APD). These are controlled by means of a field programable gate array (FPGA).}
\label{Fig:setup}
\end{figure}

The setup employed to implement and test our proposal is depicted in Fig.~\ref{Fig:setup}. The state preparation stage has a continuous-wave laser operated at 690 nm and an acousto-optic modulator (AOM). This is used to generate controlled optical pulses. Optical attenuators (not shown for sake of clarity) decrease the mean photon number per pulse to the single-photon level. Thereafter, single photons are sent through two transmissive spatial light modulators (SLM1 and SLM2), each one formed by two polarisers, quarter wave plates (QWP) and a liquid crystal display (LCD). SLM1 and SLM2 work in amplitude-only and phase-only modulation configuration, respectively. SLM2 is placed on the image plane of SLM1, and the 8-dimensional quantum system is generated with 8 parallel slits addressed on the SLMs. After SLM2 the non-normalised state of the transmitted photon is given by $|\Psi\rangle=\sum_{l=-7/2}^{7/2}\sqrt{t_l}e^{i\phi_l}|l\rangle$ \cite{LeoSlit,GlimaSlit} where $|l\rangle$ represents the state of the single photon crossing the $l$th-slit. Here, $t_l$ is the transmission for each slit controlled by SLM1; $\phi_l$ is the phase of each slit addressed by SLM2, and $N$ is a normalisation constant. The different values of $t_l$ and $\phi_l$ are configured by the grey level of the pixels in the SLMs \cite{GlimaSlit}. We addressed in the SLMs slits with the width of 2 pixels, and 1 pixel of separation between them, where each pixel is a square of 32 $\mu m$ of side length.Since a 3-qubit system is a 8-dimensional quantum system, state $|\Psi\rangle$ can be used to simulate a $3$-qubit system \cite{FastEnt,Lima14}. We generated the state $|\Psi_{U}\rangle=\frac{1}{\sqrt{8}}[1,1,1,1,1,1,1,1]$ with equal real probability amplitudes, the state $|\Psi_{GHZ}\rangle=\frac{1}{2}[1,0,0,-1,0,1,1,0]$ analogous (up to LOCC \cite{LOCC}) to the GHZ state \cite{Greenberger} and the W state $|\Psi_{W}\rangle=\frac{1}{\sqrt{3}}[0,1,1,0,1,0,0,0]$ \cite{Dur}.

\begin{figure*}[ht]
\centering
\includegraphics[width=0.95\textwidth]{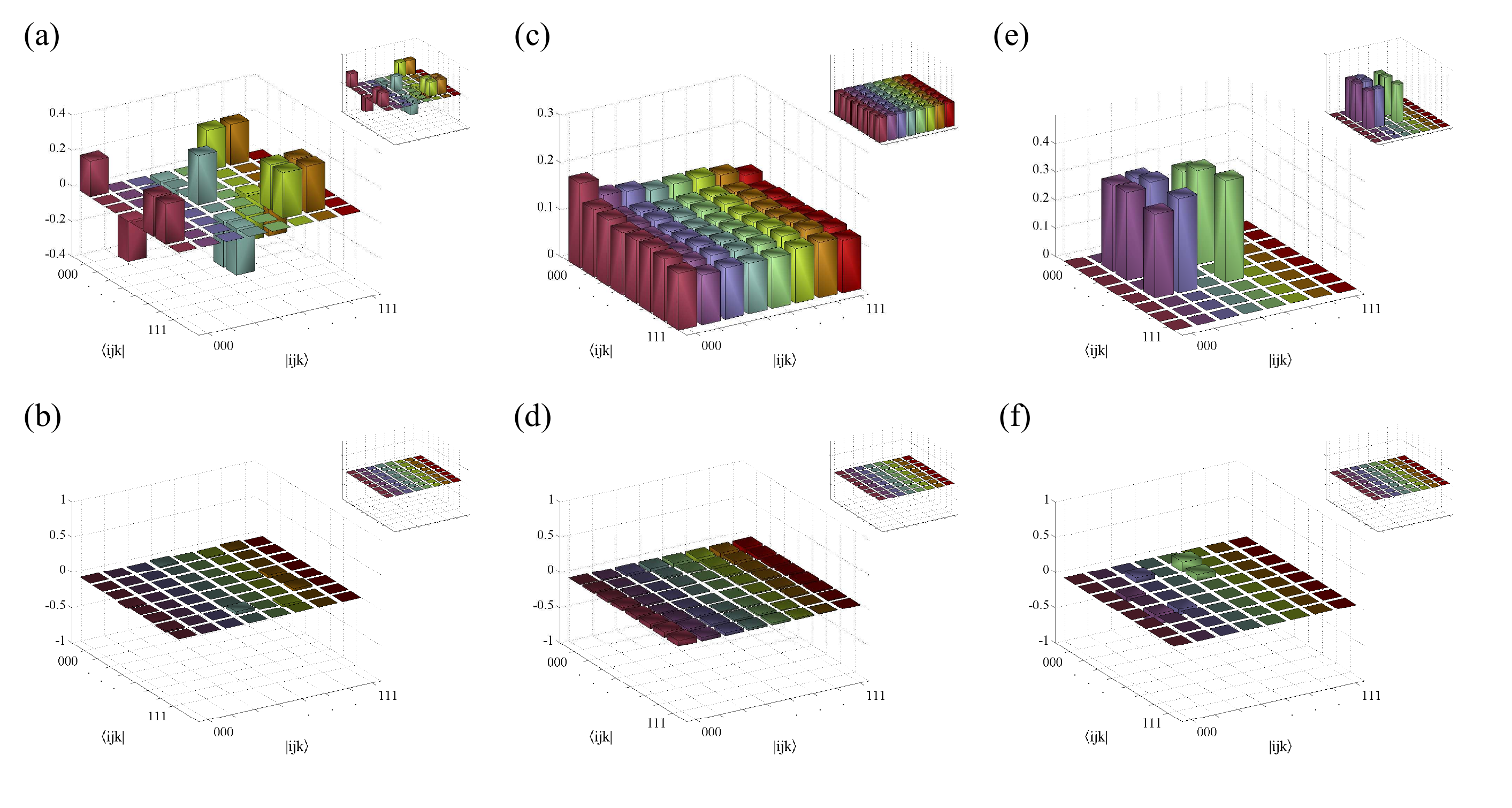}
\caption{Tomographic reconstruction for each initial spatial qudits states. Figures (a), (c) and (e) show the real parts for the reconstructed states, $|\Psi_{GHZ}\rangle$, $|\Psi_{U}\rangle$ and $|\Psi_{W}\rangle$, respectively. Imaginary parts are depicted in (b), (d) and (f). Each inset corresponds to the real and imaginary parts of the expected theoretical probabilities for the reconstruction of the initial states.  \label{Fig:a}}
\end{figure*}

To guarantee the purity of the spatial qudit states, it is necessary to observe a high visibility in the interference patterns in the far-field plane of the SLMs \cite{Tomubs}. The value of $\lambda$ can be obtained from the relation $V=(4-4\lambda)/(4-3\lambda)$, where $V$ is the observed visibility in the far-field plane. In our experiment we have $V=0.99\pm0.009$ leading to $\lambda=0.037\pm0.033$. Thus the generated states have a purity of $0.93\pm0.05$.

In order to realise the projections onto the states of the $5$ bases $\mathcal{B}_i$ we use two additional modulators, SLM3 and SLM4, working in amplitude-only and phase-only, respectively, and a point-like avalanche photo-detector (APD) \cite{Tomubs,Lima13, MinKS}. The SLMs in the projective measurement stage (see Fig. 1) are addressed with slits whose amplitudes and phases are defined to implement the projections required by our method. At the detection plane, the single-photon detection rate is proportional to the probability of projecting the initial state ($|\Psi_{GHZ}\rangle$, $|\Psi_{U}\rangle$ and $|\Psi_{W}\rangle$) into the required basis states of Eq.~(\ref{BASE1}) \cite{Tomubs,Lima13}.

From the experimental data we calculated the probability distributions associated to the 5 bases $\mathcal{B}_i$ for the three initial states. With these probability distributions and using Eq. (\ref{PHASES}), for the appropriate set of $5$ bases, we obtained a set of vectors, which were then normalised to obtain the final reconstructed states. The fidelity $F_i=|\langle \Psi_i|\Psi_{theo}\rangle|$ of the initial states with respect to the expected ones are $F_{GHZ}=0.985\pm0.015$, $F_{U}=0.96\pm0.03$ and $F_{W}=0.96\pm0.03$. These were calculated by considering the effect of the poisson noise in the detection process and selecting the highest and smallest fidelity between the expected state and the estimated state for a particular noisy set of distributions. For comparison purposes we consider an experiment with a similar configuration \cite{Tomubs} where two pure states with non vanishing, real coefficients in $d=8$ were reconstructed by means of measurements on mutually unbiased bases achieving $F=0.91\pm0.03$ and $F=0.92\pm0.03$. Note that we achieved higher fidelities with a total of $40$ projective measurements instead of $72$, as in the compared case. The states reconstructed with our method are shown in Fig.~\ref{Fig:a}. Fig.~\ref{Fig:a}(a), \ref{Fig:a}(c) and \ref{Fig:a}(e) exhibits real parts of reconstructed density matrices, compared with the expected ones (insets). Fig.~\ref{Fig:a}(b), \ref{Fig:a}(d) and \ref{Fig:a}(f) show imaginary parts of the respective matrices, compared with the theoretical predictions (insets).

We have shown that projections onto the eigenstates of a fixed set of $4$ rank-$d$ observables characterise all pure states of a $d$-dimensional quantum system up to a null measure set of dimension $d-2$. These $4d$ measurements compares favourably with the typical $d^2$ scaling of measurements of known generic tomographic methods. The addition of a fifth observable allows us to detect whether a state belongs to the null measure set or not. If this is the case, then it is always possible to construct a new set of $4$ observables which determines unambiguously the state. All sets of five observables allow us to certify whether the initial assumption on the purity of the state holds or not. We experimentally demonstrated the feasibility of our scheme in the characterisation of states in $d=8$. We achieved high fidelities by means of a reduced number of measurement outcomes and a simple post-processing method. 

\section{Acknowledgments}

We kindly acknowledge to A. Cabello, M. Grassl, P. Horodecki, R. Horodecki, S. Weigert, and K. {\.Z}yczkowski for fruitful discussions. This work was supported by Grants CONICyT PFB-0824, MSI RC130001, FONDECyT Grants N$^{\text{\underline{o}}}$ 1140636, N$^{\text{\underline{o}}}$1120067, and N$^{\text{\underline{o}}}$ 3120066. G. C. and E. S. G. acknowledge financial support of CONICyT.


\begin{thebibliography}{99}
\bibitem{Note} We generated a pure state according to the Haar measure and assumed the noise introduced by Eq. (\ref{noise-generation}). Thereafter, the five probability distributions of this state were calculated. To these probabilities we added the poisson noise characteristic of the photo-detection process and applied our method to generate an estimate $|\Psi_{est}\rangle$ of the pure state $|\Psi\rangle$. This estimate is obtained by normalising the vector provided by Eq. (\ref{solution}). Finally, the fidelity $F=|\langle\Psi|\Psi_{est}\rangle|$ between these two states was calculated and averaged over all the generated states. We considered poisson noise and $\lambda=0.03$. This latter value is characteristic of our experimental setup.
\bibitem{LOCC} The state $|\Psi_{GHZ}\rangle=\frac{1}{2}[1,0,0,-1,0,1,1,0]=(|000\rangle-|011\rangle+|101\rangle+|110\rangle)/2$ is locally equivalent to the state $(|RRR\rangle+|LLL\rangle)/\sqrt{2}$, with $|R\rangle=(|0\rangle+i|1\rangle)/\sqrt{2}$ and $|L\rangle=(|0\rangle-i|1\rangle)/\sqrt{2}$, under the transformation $\sigma_x^{(2)}\sigma_x^{(3)}$.
\end{thebibliography}
\end{document}